\def\lesssim{\mathrel{\hbox{\rlap{\hbox{\lower4pt\hbox{$\sim$}}}\hbox{$<$}}}}
\def\gtrsim{\mathrel{\hbox{\rlap{\hbox{\lower4pt\hbox{$\sim$}}}\hbox{$>$}}}}
\documentclass{aa}
\usepackage[dvips]{graphicx}
\usepackage{psfig}
\usepackage{ltxtable}
\begin{document}
\titlerunning{Abundances of nitrogen--containing molecules}
\title{The abundances of nitrogen--containing molecules during pre--protostellar collapse}
\author{D.R. Flower\inst{1}
\and G. Pineau des For\^{e}ts\inst{2,3}
\and C.M. Walmsley\inst{4}}
\institute{Physics Department, The University,
           Durham DH1 3LE, UK
\and       IAS, Universit\'{e} de Paris--Sud, F-91405 Orsay, France
\and       LUTH, Observatoire de Paris, F-92195, Meudon Cedex, France
\and       INAF, Osservatorio Astrofisico di Arcetri,
           Largo Enrico Fermi 5, I-50125 Firenze, Italy}
\offprints{C.M. Walmsley}
\abstract{}{We have studied
 the chemistry of nitrogen--bearing species during the initial stages 
of protostellar collapse, with a view to explaining the observed
 longevity of N$_2$H$^+$ and NH$_3$ and the high levels of deuteration
 of these species.}
{We followed the chemical evolution of a medium comprising
 gas and dust as it underwent free--fall gravitational collapse.
 Chemical processes which determine the relative populations of the nuclear spin states of molecules and molecular ions were included explicitly,
 as were reactions which lead ultimately to the deuteration of the 
nitrogen--containing species N$_2$H$^+$ and NH$_3$.
 The freeze--out of `heavy' molecules on to dust grains was
 taken into account.}
{We found that the timescale required for the
 nitrogen--containing species to attain their steady--state values was
 much larger than the free-fall time and
 even comparable with the probable lifetime of the precursor molecular cloud.
 However, it transpires that the chemical evolution of the gas during gravitational collapse is insensitive to its initial composition.}
 {If we suppose that the grain--sticking probabilities of atomic nitrogen and atomic oxygen are both less than unity ($S \lesssim 0.3$), we find that the observed differential freeze--out of nitrogen-- and carbon--bearing species can be reproduced by the model of free--fall collapse when a sufficiently large grain radius ($a_{\rm g} \approx 0.50$~$\mu $m) is adopted.
 Furthermore, the results of our collapse model are consistent with the high levels of deuteration of N$_2$H$^+$ and NH$_3$ which have been observed in L1544, for example, providing that $0.5 \lesssim a_{\rm g} \lesssim 1.0$~$\mu $m.
 We note that the ortho:para H$_2$D$^+$ ratio, and fractional abundance of
 ortho--H$_2$D$^+$, which is the observed form of H$_2$D$^+$, should be largest where ND$_3$ is most abundant.
\keywords{molecular cloud -- depletion  -- dust -- star formation}}

\maketitle

%
\section{Introduction}
Recent observational and theoretical studies of the initial stages of
 low--mass star formation have yielded some remarkable and intriguing results. 
The kinetic temperatures of the protostellar envelopes are observed to
be very low and, as a consequence, atomic and molecular species 
containing elements heavier than helium ultimately freeze on to
 the grain surfaces (Bacmann et al. 2002, Tafalla et al. 2004).
 As a result of this process, there remains a gas composed of atoms,
 molecules and molecular ions comprising H, D and He;
 this medium is, in effect, ``completely depleted'' of heavy elements.

Observations show that essentially all C--containing species are depleted
in the regions where CO appears to be absent (Tafalla et al. 2006);
 this is to be expected, as CO is the main gas--phase repository of carbon.
 The dissociative ionization of CO by He$^+$ produces C$^+$, and 
reactions involving C$^+$ yield C--containing molecules.
  Thus, it is very likely that the freeze--out of CO is accompanied by
 the depletion not only of directly related species, such as HCO$^+$,
 but also of molecules such as HCN, H$_2$CO and HC$_3$N.
 On the other hand, N-- or O--containing species such as OH, H$_2$O, NH$_3$
 and N$_2$H$^+$ are linked to the most abundant gas--phase forms of nitrogen and oxygen.

In the case of nitrogen, it has long seemed likely that either N or N$_2$
 is the predominant gas--phase form;
 which of these species is more abundant is uncertain, in part because
 the direct observation of either in dense clouds is not possible at
 the present time. Furthermore, chemical models are bedevilled by
 large uncertainties in the rate coefficients, at low temperatures,
 for several critical reactions, as will be mentioned below
 (Section~\ref{Nchem}). Nonetheless, it is clear that, in a situation
 where the freeze--out of species such as CO is occurring, the
 sticking probabilities (to grains) of N and N$_2$ can become crucial for the nitrogen chemistry. 

It is observed that the depletion of the heavy elements on to the grains
 occurs differentially: the nitrogen--bearing species N$_2$H$^+$ and NH$_3$
 remain in the gas phase longer than molecules such as CO
 (Bergin \& Langer 1997, Belloche \& Andr\'{e} 2004).
 In the context of a contracting cloud, ``longer'' implies higher densities.
 Tafalla et al. (2002) found that the fractional abundances of NH$_3$ and N$_2$H$^+$ in several prestellar cores either remain constant or increase with increasing density (see also Di Francesco et al. 2006). The relative longevity of NH$_3$ and N$_2$H$^+$ was believed to be due to a low binding energy of molecular nitrogen (their precursor) to the grains (Bergin \& Langer 1997, Aikawa et al. 2001). However, the recent laboratory work of Bisschop et al. (2006) has shown that the adsorption energies of N$_2$ and CO are very similar. Furthermore, the probability that N$_2$ sticks when it collides with the surface of a grain at low temperature was measured and found to be close to unity. Thus, the longevity of N$_2$H$^+$ is related neither to a low binding energy nor to a low sticking probability for N$_2$, which may be expected to freeze out similarly to CO. In this case, the ``volatile'' nitrogen is most likely to be N.

Another effect which is related to the depletion of heavy elements,
 owing to their adsorption by dust grain grains, is the enhancement
 of the abundances of the deuterated forms of those species
 remaining in the gas phase (see, for example, Ceccarelli et al. 2006).
 Given the relatively high abundances of N--containing species
 in prestellar cores, it is perhaps not surprising that the deuterated
 forms of ammonia are much more abundant than anticipated on the basis of
 the elemental $n_{\rm D}/n_{\rm H}$ ratio (cf. Roueff et al. 2005).
 In the present paper, we interpret these observations in the contexti
 of a model of a core collapsing on the free--fall timescale.
 In order to reproduce the observations, it is necessary that the
 timescale for deuterium fractionation should be comparable to both
 the dynamical timescale and the timescale for depletion on to dust grains;
 this requires $(n_{\rm g}/n_{\rm H})\pi a_{\rm g}^2$, where
 $n_{\rm g}$ is the number density and $a_{\rm g}$ is the radius of the
 grains, to be an order of magnitude smaller than is observed in
 the diffuse interstellar medium (Mathis et al. 1977), or, equivalently, the mean radius of the grains to be an order of magnitude larger.

In Section 2 we discuss the relevant chemistry and, in an Appendix,
 the treatment of the ortho and para forms of species such as NH$_3$.
 Section 3 contains our results and their comparison with the
 relevant observations, particularly of the prestellar core L1544. In Section 4, we make our concluding remarks.

\section{Chemistry}

\subsection{Nitrogen--containing species}
\label{Nchem}

The nitrogen--containing species which are observed in prestellar cores are N$_2$H$^+$ and NH$_3$. However, elemental nitrogen is expected to be mainly in the form of N and N$_2$, in the gas phase. The principal reactions converting N to N$_2$ are 

\begin{equation}
{\rm N} + {\rm OH} \rightarrow {\rm NO} + {\rm H}
\label{equ0.1}
\end{equation}
and

\begin{equation}
{\rm N} + {\rm NO} \rightarrow {\rm N}_2 + {\rm O}
\label{equ0.2}
\end{equation}
(Pineau des For\^{e}ts et al. 1990). Subsequent reactions of N$_2$ (i) with H$_3^+$ yield N$_2$H$^+$ and (ii) with He$^+$ yield N$^+$, which initiates a series of hydrogen--abstraction reactions with H$_2$, ending with NH$_4^+$, which dissociatively recombines with electrons, yielding NH$_3$. 

Pineau des For\^{e}ts et al. (1990) assumed that the reactions~(\ref{equ0.1}) and (\ref{equ0.2}) of N with OH and NO with N had small (50~K) barriers. At the kinetic temperature considered here ($T = 10$~K), the barriers contribute to making these reactions very slow: using the equilibrium fractional abundances of OH or NO, the timescale characterizing the conversion of N to N$_2$ is in excess of $10^8$~yr in molecular gas of density $n_{\rm H} = 10^4$~cm$^{-3}$. Under these circumstances, the assumption of static equilibrium, when determining the initial composition of the gas, is unlikely to be valid in so far as N$_2$ and the other products of the nitrogen chemistry are concerned. We shall return to this point in Section~\ref{Initial}.

To date, there has been no independent experimental or theoretical verification of the existence of barriers to the reactions (\ref{equ0.1}) and (\ref{equ0.2}) of N with OH and NO. Furthermore, as these are `radical--radical' reactions (i.e. reactions between two open--shell species), it is possible that there are no such barriers (Smith 1988). Accordingly, in the present study, we have removed the barriers to these reactions; at $T = 10$~K, this choice has the consequence of enhancing their rate coefficients, and reducing the corresponding timescales, by a factor of 150, to the order of $10^6$~yr; the assumption that the initial abundances are those computed in steady state then becomes more reasonable. However, these reactions remain slow in the context of the free--fall collapse, considered below, for which the timescale is $4.3\times 10^5$~yr when $n_{\rm H} = 10^4$~cm$^{-3}$ initially. It follows that simulations of  the collapse of the core by means of calculations which assume that equilibrium is attained at each (increasing) value of $n_{\rm H}$ will not predict the evolution of the nitrogen chemistry correctly.

Molecular nitrogen (N$_2$) is destroyed in the reactions 

\begin{equation}
{\rm N}_2 + {\rm H}_3^+ \rightarrow {\rm N}_2{\rm H}^+ + {\rm H}_2
\label{equ0.3}
\end{equation}

\begin{equation}
{\rm N}_2 + {\rm He}^+ \rightarrow {\rm N}_2^+ + {\rm He}
\label{equ0.4}
\end{equation}
and

\begin{equation}
{\rm N}_2 + {\rm He}^+ \rightarrow {\rm N}^+ + {\rm N} + {\rm He}
\label{equ0.5}
\end{equation}
The rate coefficients for these reactions are comparable and are of the order of the Langevin rate coefficient ($10^{-9}$~cm$^3$ s$^{-1}$). In static equilibrium at $n_{\rm H} = 10^4$~cm$^{-3}$, the fractional abundance of H$_3^+$, $n({\rm H}_3^+)/n_{\rm H} \approx 2\times 10^{-9}$, is an order of magnitude larger than the fractional abundance of He$^+$, $n({\rm He}^+)/n_{\rm H} \approx 2\times 10^{-10}$. Thus, the timescale for destroying N$_2$ is initially of the order of $10^6$~yr, comparable with the timescale for its formation and also with the timescale for its freeze--out. The total fractional abundance of H$_3^+$ and its deuterated forms and the fractional abundance of He$^+$ remain approximately constant during the free--fall collapse, up to a density $n_{\rm H} = 10^7$~cm$^{-3}$. On the other hand, the fractional abundance of OH falls rapidly, owing to freeze--out, for $n_{\rm H} \gtrsim 10^5$~cm$^{-3}$. It follows that N$_2$ is removed not only by freeze--out but also by gas--phase chemical reactions during the early phases of collapse, when it ceases to be formed. As N$_2$ is the precursor of both N$_2$H$^+$ and NH$_3$, the abundances of the latter species during collapse are dependent on the initial abundance of N$_2$.

The main process of destruction of N$_2$H$^+$ is dissociative recombination:

\begin{equation}
{\rm N}_2{\rm H}^+ + {\rm e}^- \rightarrow {\rm N}_2 + {\rm H}
\label{equ0.6}
\end{equation}

\begin{equation}
{\rm N}_2{\rm H}^+ + {\rm e}^- \rightarrow {\rm NH} + {\rm N}
\label{equ0.7}
\end{equation}
Recent ion--storage--ring measurements (Geppert et al. 2004) have shown that, somewhat counter--intuitively, the branching ratio to the products of reaction~(\ref{equ0.6}) is 0.36, smaller than the branching ratio of 0.64 to the products of reaction~(\ref{equ0.7}). We have adopted the experimental results of Geppert et al., including their expression for the total dissociative recombination coefficient [summed over the products of (\ref{equ0.6}) and (\ref{equ0.7})], namely $1.0\times 10^{-7} (T_{\rm e}/300)^{-0.51}$ cm$^3$ s$^{-1}$, where $T_{\rm e}$ is the electron temperature.

\subsection{Oxygen--containing species}
\label{Oxygen}

Specific mention should be made also of some of the uncertainties in the oxygen chemistry. Neither H$_2$O nor OH is readily observable in cold dense cores: H$_2$O lacks easily excitable emission lines, and the the 18~cm lines of OH can presently be observed only at angular resolutions such that any emission from the core is swamped by that from a less dense but warmer envelope.  Consequently, we have no direct observational evidence of the survival of species such as OH, H$_2$O and H$_3$O$^+$ in the densest regions ($n_{\rm H} > 10^5$ cm$^{-3}$) of prestellar cores. However, in somewhat less dense regions, SWAS data indicate that neither water nor molecular oxygen is a major repository of oxygen in the gas phase (Bergin et al. 2000; see also Lee et al. 2004). These authors conclude also that the best agreement with the SWAS observations is obtained if the gas--phase C:O elemental abundance ratio is greater than 0.9, in which case CO contains most of the gas--phase oxygen. This situation could arise if oxygen is mainly in the form of water ice at densities for which CO is still in the gas phase.  We have borne this result in mind when specifying the initial conditions of some of our models (Section~\ref{Initial}). 

\subsection{Deuteration reactions}
\label{Deuteration}

Under conditions of complete heavy element depletion, the principal molecule is H$_2$, and He remains in atomic form. The main ions are H$^+$ and H$_3^+$, together with its deuterated forms H$_2$D$^+$, D$_2$H$^+$ and D$_3^+$ (Roberts et al. 2003, 2004; Roueff et al. 2005). Both H$_2$D$^+$ and D$_2$H$^+$ have been observed in prestellar cores (Caselli et al. 2003, Vastel et al. 2004); D$_3^+$ is more difficult to observe, owing to the absence of a permanent dipole moment. The remarkably high levels of deuteration of H$_3^+$ which are observed arise from the differences in the zero--point vibrational energies of ions in the deuteration sequence, for example, between H$_3^+$ and H$_2$D$^+$, which has a larger reduced mass. At low kinetic temperatures, the corresponding deuteration reaction 

\begin{equation}
{\rm H}_3^+ + {\rm HD} \rightleftharpoons {\rm H}_2{\rm D}^+ + {\rm H}_2
\label{equ1}
\end{equation}
is strongly favoured, energetically, in the forwards direction, when ground state reactants and products are involved. The other reactant, HD, is formed, like H$_2$, on the surfaces of dust grains.

In earlier papers (Walmsley et al. 2004, Flower et al. 2004, 2005, 2006), we considered the deuteration of H$_3^+$ and emphasized the importance of distinguishing between the ortho and para forms of the reactants and the products. Different nuclear spin states are, through the appropriate quantum statistics (Fermi--Dirac or Bose--Einstein), associated with different rotational states of the molecule. In the best--known example of H$_2$, ortho states (with total nuclear spin $I = 1$) are associated with rotational levels with odd values of $J$, whereas para states ($I = 0$) are associated with even values of $J$. Thus, the lowest ortho state, $J = 1$, lies approximately 170~K higher than the lowest para state, $J = 0$, which is the ground state of the molecule. This energy, which is large compared with the thermal energy of the gas (the kinetic temperature $T \approx 10$~K) becomes available in the backwards reaction~(\ref{equ1}), in the case where {\it ortho}--H$_2$ is the reactant with H$_2$D$^+$. As the backwards reaction reverses the deuteration of H$_3^+$, it is essential to know the ortho:para H$_2$ ratio in order to predict correctly the degree of deuteration (Gerlich et al. 2002, Flower et al. 2006).

It is well established that H$_3^+$ plays a pivotal role in interstellar chemistry. Similarly, its deuterated forms, H$_2$D$^+$, D$_2$H$^+$ and D$_3^+$, are determinants of the degree of deuteration of other species. The reaction

\begin{equation}
{\rm N}_2 + {\rm H}_2{\rm D}^+ \rightleftharpoons {\rm N}_2{\rm D}^+ + {\rm H}_2
\label{equ2}
\end{equation}
and the analogous reactions with D$_2$H$^+$ and D$_3^+$ are the keys to understanding the N$_2$D$^+$:N$_2$H$^+$ ratio in prestellar cores. The degree of deuteration of NH$_3$ is determined by analogous reactions with the deuterated forms of H$_3^+$. Thus, the interpretation of the (high) levels of deuteration of both N$_2$H$^+$ and NH$_3$ observed in prestellar cores involves consideration of the ortho:para abundance ratios of certain gas--phase species. 

\section{Results}
\label{Results}

We have considered the gravitational contraction (free--fall collapse) of a condensation of gas and dust with an initial number density, $n_{\rm H} = n({\rm H}) + 2n({\rm H}_2)$, in the range $10^3 \le n_{\rm H} \le 10^5$~cm$^{-3}$ and kinetic temperature $T = 10$~K. The initial value of the grain radius was varied in the range $0.05 \le a_{\rm g} \le 1.0$~$\mu $m, for reasons that will become apparent below. The smallest value, $a_{\rm g} = 0.05$~$\mu $m, yields approximately the same grain opacity, $n_{\rm g}\pi a_{\rm g}^2$, as the grain size distribution of Mathis et al. (1977), with limits of $0.01 \le a_{\rm g} \le 0.3$~$\mu $m, and where $n_{\rm g}$ is the grain number density. 

The initial composition of the gas was computed on the assumption of static equilibrium (``steady state''). The elemental abundances and initial depletions are specified in table 1 of Flower et al. (2005); they yield an initial gas--phase C:O elemental abundance ratio, $n_{\rm C}/n_{\rm O} = 0.67$, and a dust:gas mass ratio of 0.0094. In the light of the discussion in Section~\ref{Oxygen}, higher values of $n_{\rm C}/n_{\rm O}$ were also considered. The chemistry comprised reactions involving species containing H, He, C, N, O and S and distinguished between the possible nuclear spin states of H$_3^+$ and its deuterated forms, and those of H$_2^+$ and H$_2$. It also distinguished between the nuclear spin states of nitrogen--containing species, following the discussion in Appendix~\ref{AppA}. The chemistry and list of species are available from http://massey.dur.ac.uk/drf/protostellar/chemistry\_species.

In order to study the deuteration of nitrogen--containing molecules, such as NH$_3$, and ions, such as N$_2$H$^+$, we included the relevant chemical reactions but did not  introduce explicitly the different nuclear spin states of the {\it deuterated} forms. Thus, NH$_2$D, for example, was treated as a single chemical species; we did not distinguish between its ortho and para forms. This simplification was necessary to ensure that the chemistry remained manageable, but it did not, in our view, compromise the reliability of the results relating to the degrees of deuteration of nitrogen--containing species.

In most of the models, the grain sticking probability $S({\rm N}) = 0.1$ for atomic nitrogen, and $S({\rm O}) = 0.1$ also (see the discussion in Section~\ref{freefall}). Larger values of $S({\rm N})$ and $S({\rm O})$ were considered, and reference to the results of these calculations will be made below. For all other atomic and molecular species, $S = 1.0$. The rate of cosmic ray ionization of hydrogen was taken to be $\zeta = 1\times 10^{-17}$ s$^{-1}$, as in our earlier papers (Flower et al. 2005, 2006). 

\subsection{Initial abundances}
\label{Initial}

\begin{figure}
\centering
\includegraphics[height=10cm]{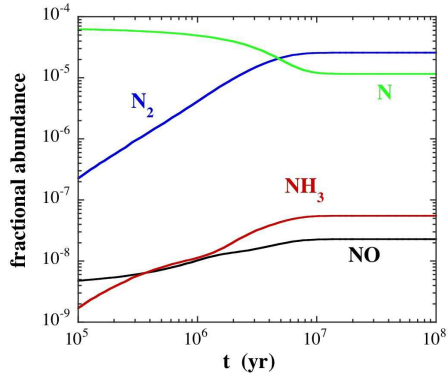}
\caption{The fractional abundances, relative to $n_{\rm H} \approx n({\rm H}) + 2n({\rm H}_2)$, of N, N$_2$, NO and NH$_3$ as functions of time, $t$, through to the attainment of their steady--state values. A constant gas density of $n_{\rm H} = 10^4$~cm$^{-3}$ and temperature $T = 10$~K were assumed and freeze--out of the heavy elements was neglected.}
\label{s-s1}
\end{figure}

As we noted in Section~\ref{Nchem}, the timescales associated with the chemistry of the nitrogen--bearing species are particularly large -- even when, as here, we assume that there are no barriers to the initiating reactions~(\ref{equ0.1}) and (\ref{equ0.2}). To illustrate this point, we plot, in Fig.~\ref{s-s1}, the fractional abundances of gas--phase N, N$_2$, NO and NH$_3$ as functions of time, $t$, at a constant gas density of $n_{\rm H} = 10^4$~cm$^{-3}$, starting with the nitrogen in atomic form. Freeze--out of the heavy elements is neglected in this calculation.

Figure~\ref{s-s1} shows that the formation of molecular nitrogen requires $t \approx 5\times 10^6$~yr; the fractional abundances of NO and NH$_3$ attain steady state on a similar timescale, as they are a precursor to and a consequence of, respectively, the formation of N$_2$. This timescale is comparable with the lifetimes, $\tau $, of molecular clouds, as derived by Tassis \& Mouschovias (2004; $\tau \approx 10^7$~yr), but greater than the ages of clouds in the solar vicinity, as determined by Hartmann et al. (2001; $\tau \approx 10^6$ yr). Thus, it seems possible, though by no means certain, that such species have reached their steady--state abundances before protostellar collapse begins. 

In fact, the nitrogen chemistry is peculiar, in that it is activated by a reaction between two neutral species (N and OH) of low fractional abundance (of the order of $10^{-5}$ and $10^{-8}$, respectively, in steady--state). Overall, the steady--state chemistry is dominated by ion--neutral reactions involving H$_2$ as one of the reactants. Although the timescale for cosmic ray ionization is very long, of the order of $\zeta ^{-1} \approx 10^9$ yr, the principal ionization processes involve H$_2$ and He, which are the most abundant species in the gas; the charge is then passed to `heavy' species, in rapid ion--neutral reactions. The resulting rate of ionization of the heavy species, X, is approximately $n_{\rm H}/n_{\rm X}$ faster than the rate of direct cosmic ray ionization of X. As a consequence, the abundances of ions and the species which form through ion--neutral reactions, such as OH, reach steady state on a timescale which is about an order of magnitude smaller than the free--fall time of $4.3\times 10^5$~yr for an initial density $n_{\rm H} = 10^4$~cm$^{-3}$.

In view of the uncertainty in the gas--phase C:O elemental abundance ratio (see Section~\ref{Oxygen}), we recalculated the steady--state composition of the gas with a higher value of this ratio, $n_{\rm C}/n_{\rm O} = 0.95$. We found that, although the fractional abundances of OH and NO decreased, by a factor of aproximately 2.5, that of N$_2$ fell by only 15\%, with similar decreases for N$_2$H$^+$ and NH$_3$. We conclude that, as long as the gas--phase C:O abundance ratio remains less than 1, there remains sufficient oxygen in the gas phase in the form of OH to drive the nitrogen chemistry. In any case, reactions such as

\begin{equation}
{\rm He}^+ + {\rm CO} \rightarrow {\rm C}^+ + {\rm O} + {\rm He}
\label{equ17}
\end{equation}
and

\begin{equation}
{\rm He}^+ + {\rm O}_2 \rightarrow {\rm O}^+ + {\rm O} + {\rm He}
\label{equ18}
\end{equation}
restore atomic oxygen to the gas phase, thereby enabling OH to form in the sequence of reactions initiated by H$_3^+$(O, H$_2$)OH$^+$. 

We note that the analogous reaction~(\ref{equ0.5}) of He$^+$ with N$_2$ not only leads ultimately to the formation of NH$_3$ but also restores atomic nitrogen to the gas phase. For this reason, the chemical evolution predicted by the model calculations proves to be insensitive to the initial value of the atomic to molecular nitrogen abundance ratio; see the results in the following Section~\ref{freefall}.

\subsection{Free--fall collapse}
\label{freefall}

The free--fall time is given by 
$$\tau _{\rm ff} = \left [ \frac {3\pi }{32G\rho _0} \right ]^{\frac {1}{2}}$$
where $\rho _0$ is the initial mass density; $\tau _{\rm ff} = 4.3\times 10^5$~yr for an initial number density $n_{\rm H} = 10^4$~cm$^{-3}$. We have assumed that the density evolves on the free--fall timescale. (We note that Ward--Thompson et al. (2006) have presented recently statistical arguments which suggest that core lifetimes are a few times $\tau _{\rm ff}$, possibly owing to magnetic support). 

The freeze--out time 
$$\tau _{\rm fo} = \left [ n_{\rm g} \pi a_{\rm g}^2 v_{\rm th} S \right ]^{-1}$$
where $n_{\rm g} \pi a_{\rm g}^2$ is the grain opacity and $v_{\rm th}$ is the thermal speed of the species which sticks to the grain, with a sticking probability $S$. For $a_{\rm g} = 0.05$~$\mu $m, a molecule such as CO freezes out on a timescale  $\tau _{\rm fo} = 2.2\times 10^5$ yr when $n_{\rm H} = 10^4$ cm$^{-3}$. Thus, the free--fall and the freeze--out times are comparable at this density and for this value of the grain radius. We note that $n_{\rm g} \propto a_{\rm g}^{-3}$ and hence $\tau _{\rm fo} \propto a_{\rm g}$: freeze--out occurs at later times and consequently higher densities if the grains are larger; this effect will be seen in the results presented below. 

\subsubsection{Dependence on the initial conditions}

\begin{figure}
\centering
\includegraphics[height=20cm]{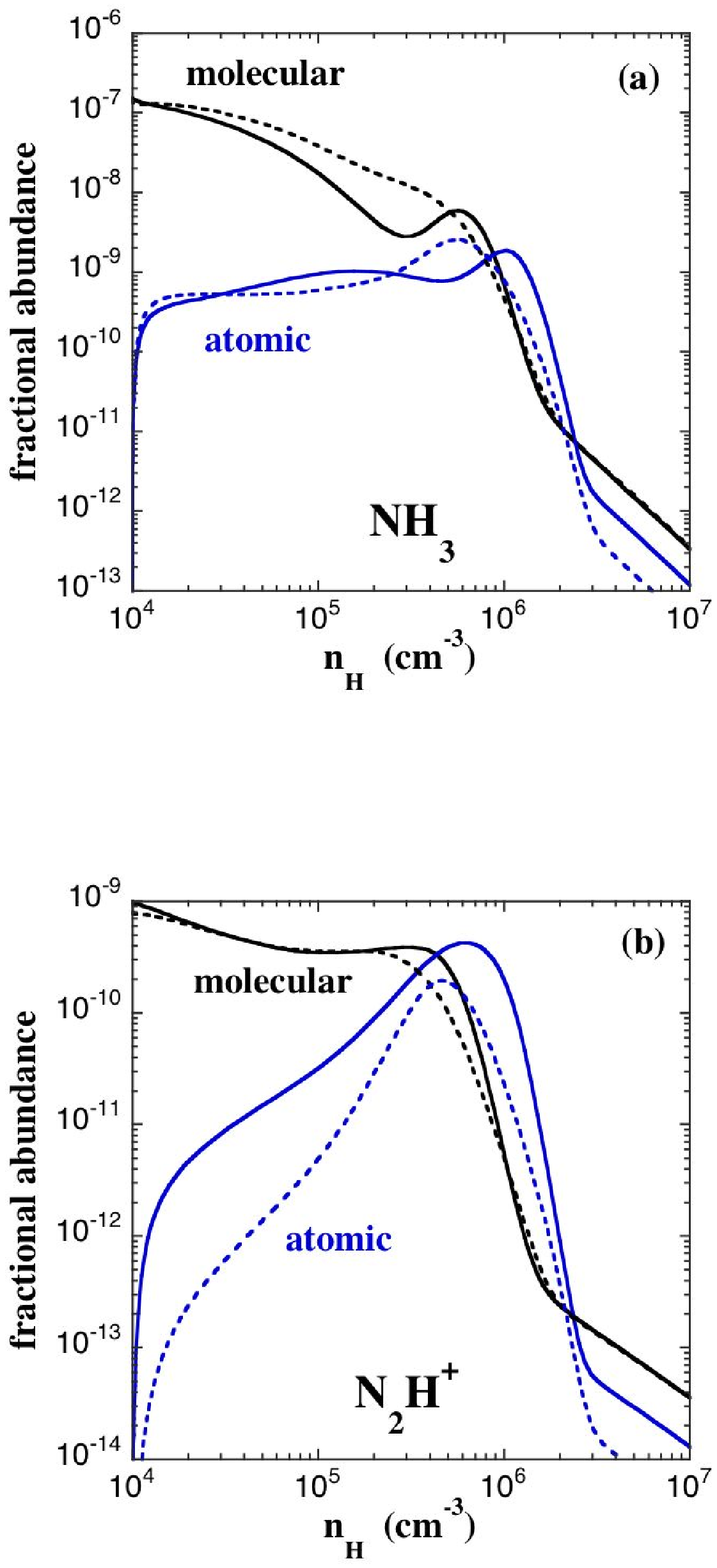}
\caption{The fractional abundances, relative to $n_{\rm H} \approx n({\rm H}) + 2n({\rm H}_2)$, of (a) NH$_3$ and (b) N$_2$H$^+$, as functions of $n_{\rm H}$, during free--fall gravitational collapse from an initial density $n_{\rm H} = 10^4$~cm$^{-3}$. A constant temperature $T = 10$~K and a constant grain radius $a_{\rm g} = 0.20$~$\mu $m were assumed. The full curves correspond to an elemental C:O ratio, in the gas phase, of $n_{\rm C}/n_{\rm O} = 0.67$ (our standard assumption), and the broken curves to $n_{\rm C}/n_{\rm O} = 1.0$. In the ``molecular'' case, the gas was initially in molecular form (our standard assumption), whereas, in the ``atomic'' case, the initial composition was atomic, with the exceptions of H$_2$, HD and CO. In these calculations, $S({\rm N}) = S({\rm O}) = 0.1$.}
\label{initcond}
\end{figure}

The uncertainties in the initial chemical composition of the gas has already been considered on Section~\ref{Initial}, and, in particular, the fact that the chemistry may not have attained steady state before gravitational collapse begins. Accordingly, we have carried out a number of test calculations, varying the intial composition from equilibrium. These calculations were designed to test the sensitivity of the fractional abundance profiles, during the subsequent free--fall collapse, to such variations. In Fig.~\ref{initcond} is shown a sample of such calculations. Fig.~\ref{initcond} enables the consequences of assuming that nitrogen is initially in atomic rather than mainly in molecular form to be appreciated. There is a large effect initially on the NH$_3$ and N$_2$H$^+$ abundances, but, by densities of a few times $10^5$ cm$^{-3}$, the differences have largely disappeared.
Furthermore, and for reasons given above, the effect of setting $n_{\rm C}/n_{\rm O} = 1.0$ initially, and thus reducing drastically the
oxygen available to form OH, is less than might have been anticipated, although the
formation of N$_2$ is slower than when $n_{\rm C}/n_{\rm O} = 0.67$.

Fig.~\ref{initcond} demonstrates the remarkable insensitivity of the chemical evolution during collapse to the initial conditions, even in the case of the nitrogen--containing species, whose chemistry involves relatively long timescales (cf. Section~\ref{Nchem}). Major variations of the initial conditions from steady--state, such as assuming that all the gas--phase nitrogen is initially in atomic form, have transient and limited effects on the subsequent evolution. In essence, Fig.~\ref{initcond} confirms the effectiveness of the ion--neutral chemistry, which we have already considered in Section~\ref{Initial}.

\subsubsection{Differential depletion on to grains}

As noted in the Introduction, an intriguing observational result, which awaits an  interpretation, is the relative longevity of N$_2$H$^+$ and NH$_3$ during freeze--out of the heavy elements from the gas phase. Explanations in terms of a low adsorption energy or a low sticking probability for N$_2$, the precursor of both N$_2$H$^+$ and NH$_3$, appear to have been invalidated by recent experimental work (Bisschop et al. 2006). However, we are unaware of any measurements of the sticking probability, $S({\rm N})$, of {\it atomic} nitrogen (or $S({\rm O})$, which is also relevant to the gas--phase chemistry; see below). N and O are lighter than N$_2$ and have correspondingly higher vibrational frequencies when interacting with the grain surface, which, in the present context, is probably composed essentially of CO, deposited during freeze--out. The issue which waits to be addressed, experimentally or theoretically, is the strength of CO -- N and CO -- O bonds, relative to CO -- CO (or CO -- N$_2$). We note that there exists some experimental evidence for the enrichment of CO, relative to N$_2$, during freeze--out at low temperatures, although on to a layer of predominantly water ice, rather than CO ice (Notesco \& Bar-Nun 1996). 

The formation of molecular nitrogen is initiated by reaction~(\ref{equ0.1}), N(OH, H)NO, which requires both N and OH to be available in the gas phase. The hydroxyl radical, OH, is formed in the dissociative recombinations of H$_2$O$^+$ and H$_3$O$^+$, which are produced from atomic oxygen, through an initial proton--transfer reaction with H$_3^+$ and subsequent hydrogenation of OH$^+$ and H$_2$O$^+$ by H$_2$. These chemical considerations suggest that relatively low sticking probabilities for both atomic nitrogen and atomic oxygen might contribute to  the longevity of N$_2$H$^+$ and NH$_3$. A low value of $S({\rm O})$ would also have consequences for the abundances of `pure' oxygen--containing species, such as O, O$_2$ and H$_2$O; but the angular resolution and sensitivity of current instrumentation are such that the core cannot be distinguished from the more extended, lower density regions that are observed, for example, in OH (Goldsmith \& Li 2005).

Although our results are not directly comparable to such observations (which have a linear spatial resolution typically in the range 0.15--0.3~pc), it is unlikely that the chemical abundances at densities of order $10^4$ cm$^{-3}$ (our initial value) differ greatly from those observed in clouds with densities of a few times $10^3$ cm$^{-3}$, which are believed to be relevant to the measurements of OH, for example. Goldsmith \& Li (2005) found $n({\rm OH})/n({\rm H}_2) = 2\times 10^{-8}$ in L1544, Bergin \& Snell (2002) obtained upper limits to $n$(ortho-H$_2$O/$n$(H$_2$) of the order of $3\times 10^{-8}$ in two clouds, and Pagani et al. (2003) derived $n({\rm O}_2)/n({\rm H}_2) \lesssim 10^{-7}$ in several sources. The observational results for O$_2$, in particular, plead in favour of models in which $n_{\rm C}/n_{\rm O} = 1.0$ initially in the gas phase: these models have steady--state O$_2$ abundances close to the observed upper limit. On the other hand, when $n_{\rm C}/n_{\rm O} = 0.67$ initially, the steady--state abundance of O$_2$ is roughly two orders of magnitude larger. However, we recall that the nitrogen chemistry is not  sensitive to the initial value of $n_{\rm C}/n_{\rm O}$.

\begin{figure}
\centering
\includegraphics[height=20cm]{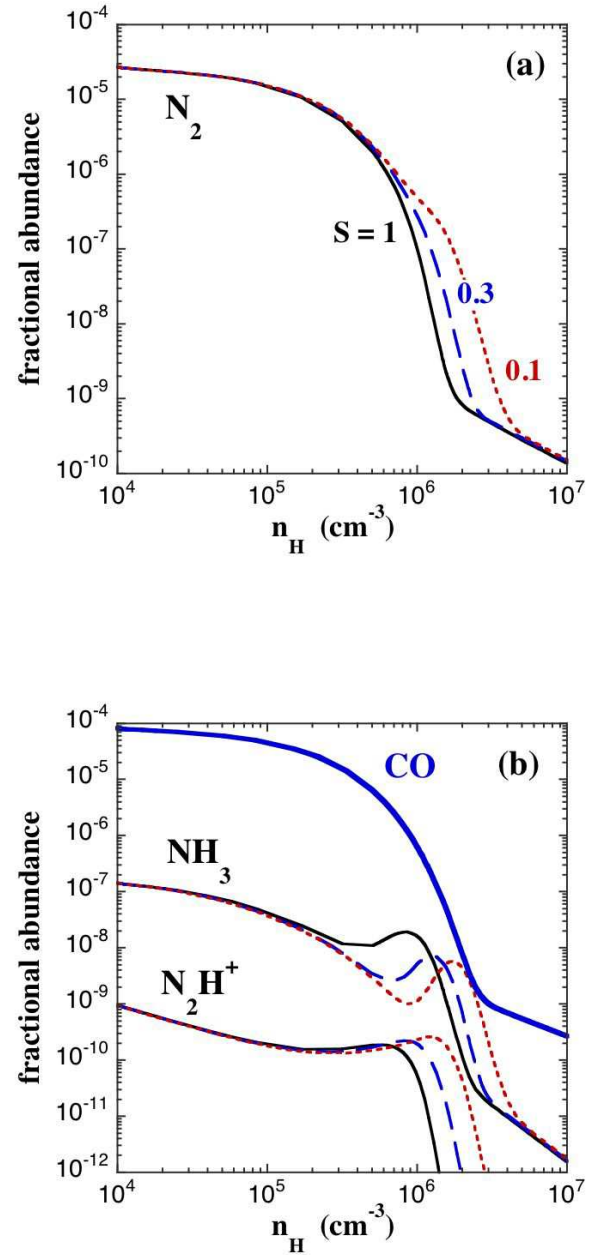}
\caption{The fractional abundances, relative to $n_{\rm H}$, of (a) N$_2$, and (b) CO, NH$_3$ and N$_2$H$^+$, as functions of $n_{\rm H}$, predicted by the free--fall model with an initial density $n_{\rm H} = 10^4$ cm$^{-3}$ and for a grain radius $a_{\rm g} = 0.5$~$\mu $m. The adopted values of the sticking probabilities $S({\rm N}) = S({\rm O})$ are indicated in panel (a), and the key is identical in panel (b).}
\label{s-s2}
\end{figure}

In Fig.~\ref{s-s2} are shown the profiles of N$_2$, NH$_3$ and N$_2$H$^+$, as functions of $n_{\rm H}$, in the course of free--fall collapse from an initial steady--state composition and an initial density $n_{\rm H} = 10^4$ cm$^{-3}$; the profile of CO is also plotted, for comparison. The values adopted for $S({\rm N}) = S({\rm O})$ are indicated in panel~(a) of Fig.~\ref{s-s2}, and $S = 1.0$ for all other atomic and molecular species. As $S({\rm N})$ and $S({\rm O})$ are reduced, atomic nitrogen and also OH, which is produced from atomic oxygen, are removed from the gas phase more slowly than CO, and the nitrogen chemistry continues to be driven by reactions~(\ref{equ0.1}) and (\ref{equ0.2}). Consequently, the abundances of the nitrogen--containing species increase, relative to CO, as the density approaches $n_{\rm H} = 10^6$ cm$^{-3}$. Supplementary calculations have confirmed that the increase in the N$_2$H$^+$:CO ratio, for example, occurs only when $S$(N) and $S$(O) are reduced simultaneously. Thus, we conclude tentatively that the longevity of N$_2$H$^+$ and NH$_3$ is related to low grain sticking probabilities for both atomic nitrogen and oxygen.

\begin{figure}
\centering
\includegraphics[height=20cm]{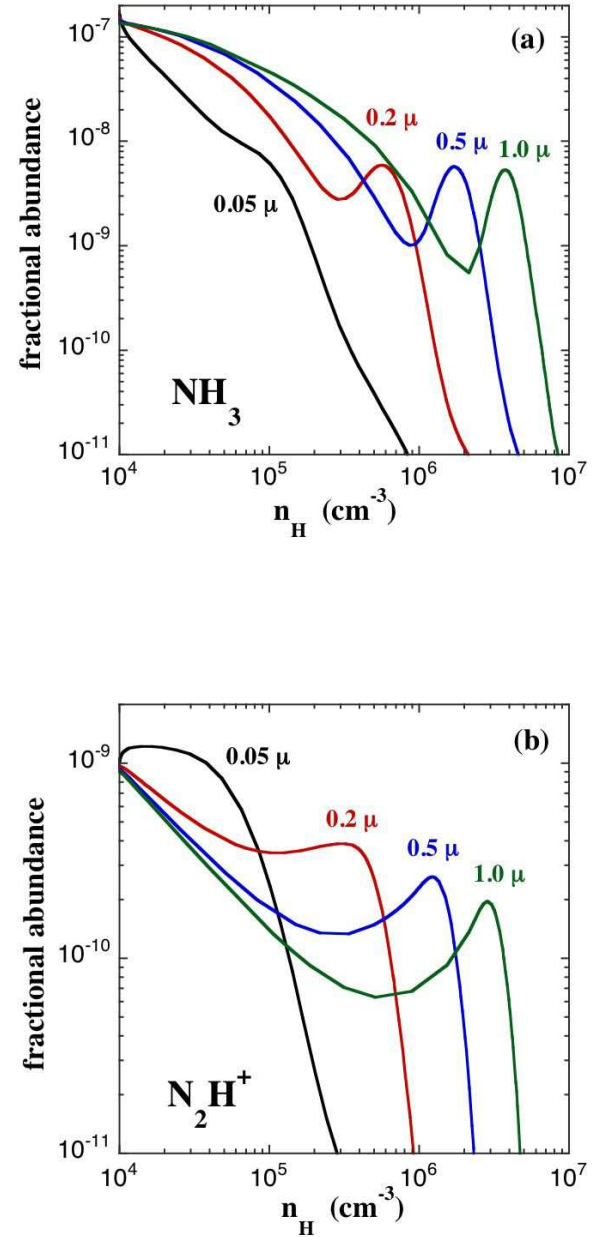}
\caption{The fractional abundances, relative to $n_{\rm H}$, of (a) NH$_3$ and (b) N$_2$H$^+$, predicted by the free--fall model with an initial density $n_{\rm H} = 10^4$ cm$^{-3}$, for grain sizes in the range $0.05 \le a_{\rm g} \le 1.0$~$\mu $m. $S({\rm N}) = S({\rm O}) = 0.1$ and $S = 1.0$ for all other atomic and molecular species.}
\label{s-s2a}
\end{figure}

We show the effect of varying the grain size in Fig.~\ref{s-s2a}. The onset of freeze--out is delayed to higher gas densities as $a_{\rm g}$ increases, owing to the reduction in the total grain surface area, which, in our model, is inversely proportional to $a_{g}$. Thus, $a_{\rm g} = 0.5$~$\mu $m corresponds to a value of $(n_{\rm g}/n_{\rm H})\pi a_{\rm g}^2 = 1.7\times 10^{-22}$~cm$^2$, which is an order of magnitude smaller than is deduced from observations of dust in the diffuse interstellar medium (see, for example, Mathis et al. 1977). CO is depleted by a factor of e by a density $n_{\rm H} \approx 6\times 10^4$~cm$^{-3}$ for $a_{g} = 0.20$~$\mu $m and $n_{\rm H} \approx 2\times 10^5$~cm$^{-3}$ for $a_{g} = 0.50$~$\mu $m. These values of $n_{\rm H}$ may be compared with the densities characterizing  CO freeze--out, deduced from observations of five nearby cores (Tafalla et al. 2002, table 4), namely $3\times 10^4 \lesssim n_{\rm H} \lesssim 1\times 10^5$~cm$^{-3}$ (note that Tafalla et al. expressed their results in terms of $n({\rm H}_2) \approx n_{\rm H}/2$). 

It is interesting that, according to our models (Fig.~\ref{s-s2}), there is a wide range of density, $1\times 10^5 \lesssim n_{\rm H} \lesssim 2\times 10^6$~cm$^{-3}$, in which the fractional abundance of N$_2$H$^+$ is roughly constant. After an initial drop, the fractional abundance of ammonia goes through a maximum at $n_{\rm H} \approx 10^6$~cm$^{-3}$; this is reminiscent of the deduction of Tafalla et al. (2002), that, in the five cores which they observed, the fractional abundance of N$_2$H$^+$ remains roughly density--independent whilst that of NH$_3$ appears to increase with increasing core density. Taking into account the uncertainties in the chemistry, we find this qualitative agreement to be encouraging.

\begin{figure}
\centering
\includegraphics[height=20cm]{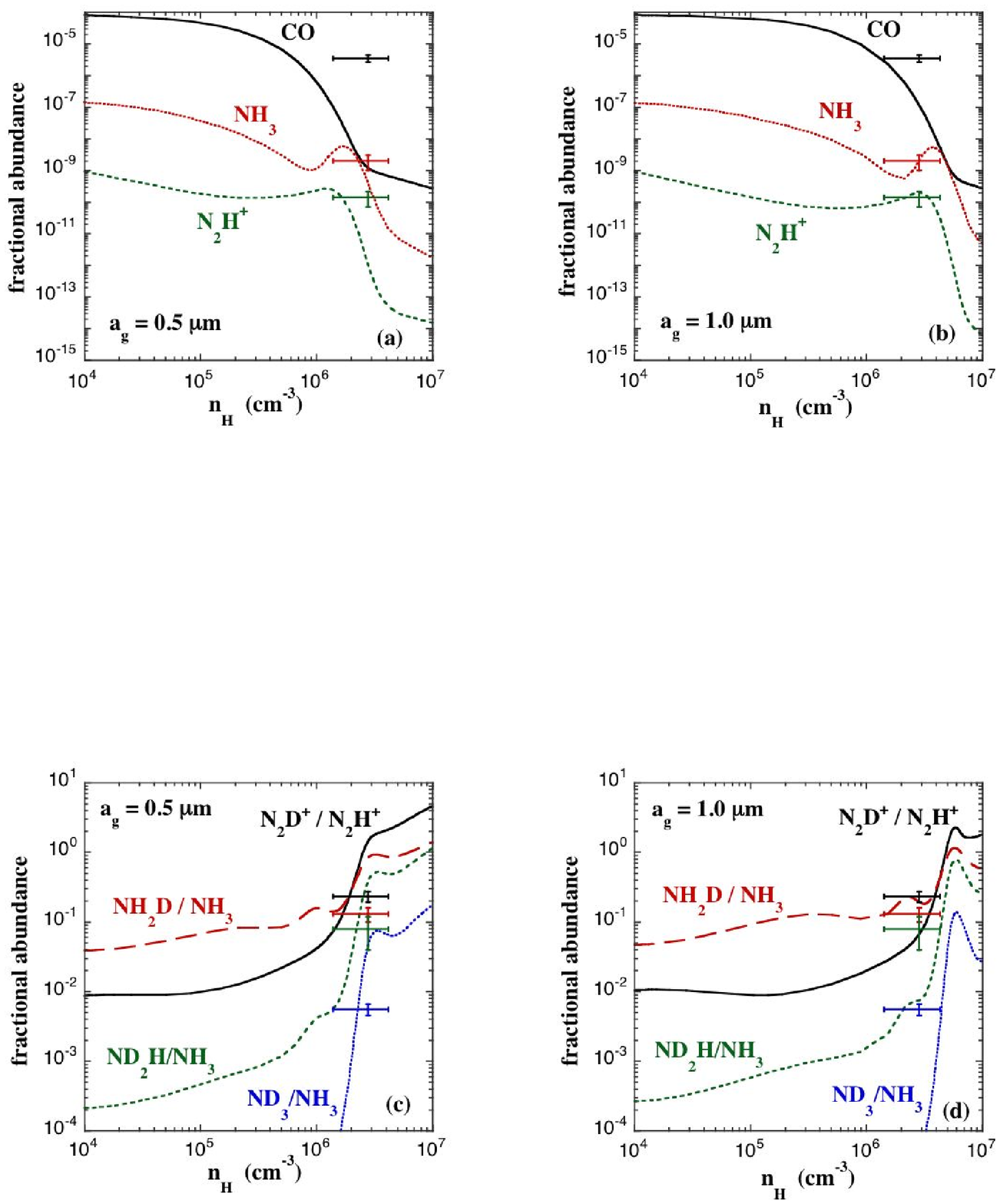}
\caption{The calculated and observed values of the fractional abundances, relative to $n_{\rm H}$, of CO, NH$_3$ and N$_2$H$^+$, and the corresponding values of the relative abundances of the deuterated forms of NH$_3$ and N$_2$H$^+$. The calculated values derive from the free--fall model with an initial density $n_{\rm H} = 10^4$~cm$^{-3}$ and grain radii $a_{\rm g} = 0.5$~$\mu $m and $a_{\rm g} = 1.0$~$\mu $m. $S({\rm N}) = S({\rm O}) = 0.1$ and $S = 1.0$ for all other atomic and molecular species. The observed ratios are denoted by the crosses, with error bars.}
\label{s-s2b}
\end{figure}

With the comments in the previous paragraphs in mind, we display, in the upper panels of Fig.~\ref{s-s2b}, the fractional abundances of CO, N$_2$H$^+$ and NH$_3$, observed in the prestellar core L1544 (Tafalla et al. 2004), along with the estimated error bars; the density, $n_{\rm H}$, has been deduced from the peak intensity of the dust emission. The abundance profiles which derive from the free--fall model, with an initial density $n_{\rm H} = 10^4$~cm$^{-3}$ and grain radii $a_{\rm g} = 0.5$~$\mu $m and $a_{\rm g} = 1.0$~$\mu $m, are also plotted in Fig.~\ref{s-s2b}. The observed fractional abundances are reproduced satisfactorily by the model with $a_{\rm g} = 1.0$~$\mu $m. As we shall see in Section~\ref{deut} below, the evidence from the calculated and observed levels of deuterium fractionation of N$_2$H$^+$ and NH$_3$ is also consistent with this value of $a_{\rm g}$.

\subsection{Deuteration of nitrogen--containing species}
\label{deut}

Observations of prestellar cores have shown that, not only are nitrogen--containing species relatively long--lived, but also they exhibit remarkably high degrees of deuteration. Indeed, even triply--deuterated ammonia, ND$_3$, has been detected (van der Tak et al. 2002, Lis et al. 2002), with an abundance, relative to NH$_3$, which is more than 10 orders of magnitude larger than would be anticipated on the basis of the $n_{\rm D}/n_{\rm H}$ ($\approx 10^{-5}$) elemental abundance ratio. 

We have already considered, in Section~\ref{Deuteration}, the mechanism of deuteration of nitrogen--containing species such as NH$_3$ and N$_2$H$^+$, which involves reactions with the deuterated forms of H$_3^+$. H$_2$D$^+$ is formed through the deuteration of H$_3^+$ by HD and (together with D$_2$H$^+$ and D$_3^+$) is the main agent of deuteration of other species. The fractional abundances of the deuterated forms of H$_3^+$ become so large that the timescale characterizing the deuteration of N$_2$H$^+$, in reaction~(\ref{equ2}) and the analogous reactions with D$_2$H$^+$ and D$_3^+$, falls to values comparable with the free--fall time. Consequently, the level of deuteration of N$_2$H$^+$ increases with the gas density, $n_{\rm H}$, in the course of the collapse.

\begin{figure}
\centering
\includegraphics[height=20cm]{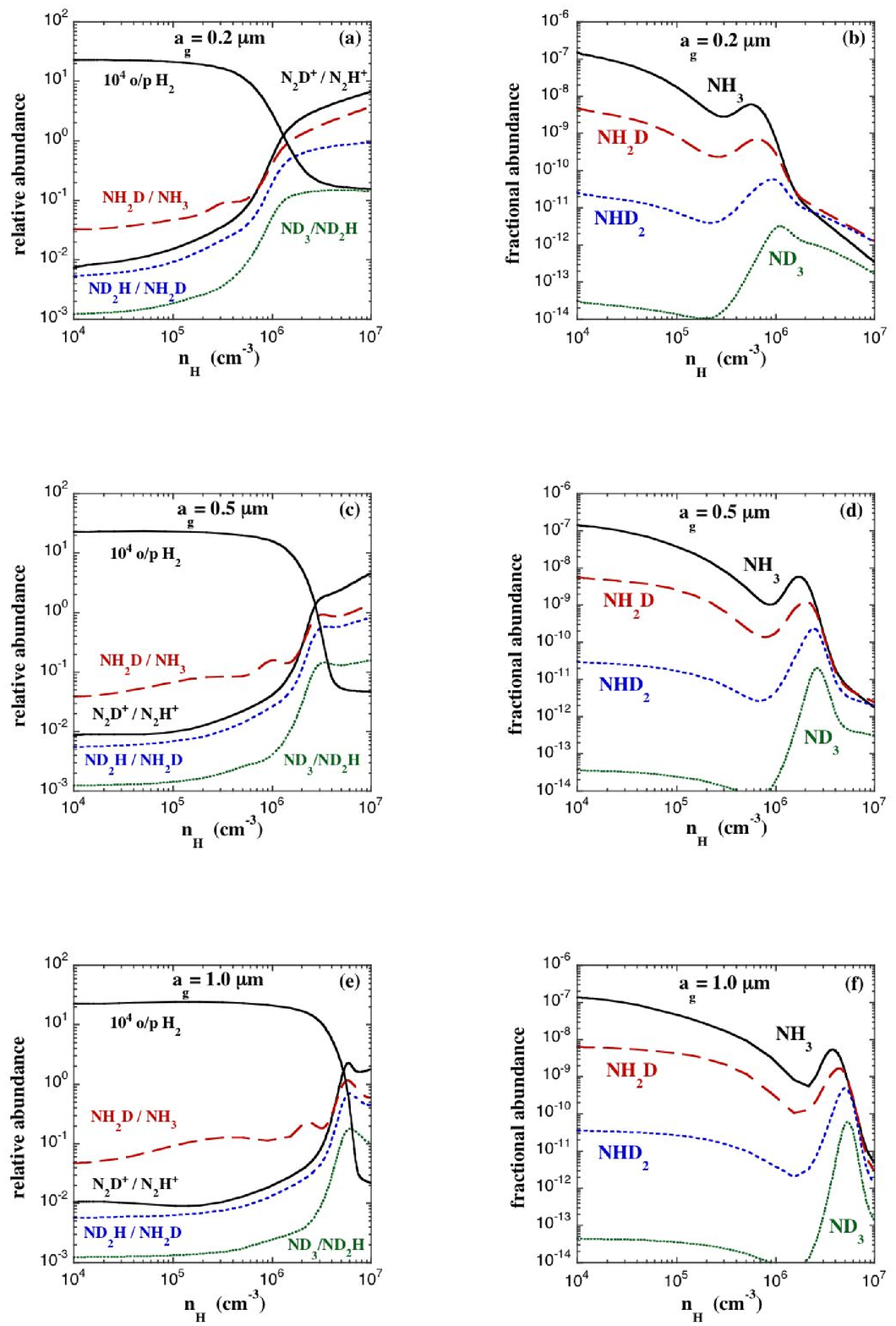}
\caption{(a, c, e) The abundance ratios of the deuterated forms of N$_2$H$^+$ and NH$_3$, together with ($10^4$ times) the ortho:para H$_2$ ratio, and (b, d, f) the fractional abundances of the deuterated forms of ammonia, in the gas phase, as functions of $n_{\rm H}$; (a) and (b) correspond to $a_{\rm g} = 0.2$~$\mu $m, (c) and (d) to $a_{\rm g} = 0.5$~$\mu $m, (e) and (f) to $a_{\rm g} = 1.0$~$\mu $m.}
\label{NHD}
\end{figure}

As shown by Roueff et al. (2005, Table~8; see also Lis et al. 2006), the population densities of successive stages of deuteration of NH$_3$ are observed to decrease by only about an order of magnitude, and $10^{-4} \lesssim n({\rm ND}_3)/n({\rm NH}_3) \lesssim 10^{-2}$ in the prestellar cores which have been observed. Similarly, N$_2$D$^+$ is observed to be approximately an order of magnitude less abundant than N$_2$H$^+$ (Caselli et al. 2002). In Fig.~\ref{NHD}, these isotopic abundance ratios are plotted, together with the fractional abundances of NH$_3$ and its deuterated forms. The fractional abundances of the deuterated forms of ammonia attain a maximum in a rather narrow range of density $5\times 10^5 \lesssim n_{\rm H} \lesssim 5\times 10^6$ cm$^{-3}$. The maximum fractional abundance shifts to higher densities with increasing degree of deuteration and with increasing grain size. Thus, ND$_3$, for example, should sample higher density gas than NH$_3$. 

Also plotted in Fig.~\ref{NHD} is the ortho:para H$_2$ ratio, in view of the key role which it plays in the deuteration process. The ortho:para H$_2$ ratio tends towards thermalization later (i.e. at higher $n_{\rm H}$) with increasing grain size. Because freeze--out occurs later for larger $a_{\rm g}$, the fractional abundances of H$_3^+$ and H$^+$ maximize later, which, in turn, delays ortho--to--para conversion in proton exchange reactions of these species with H$_2$; this has knock--on effects on the ortho:para ratios of other species, such as H$_2$D$^+$. 

\begin{figure}
\centering
\includegraphics[height=20cm]{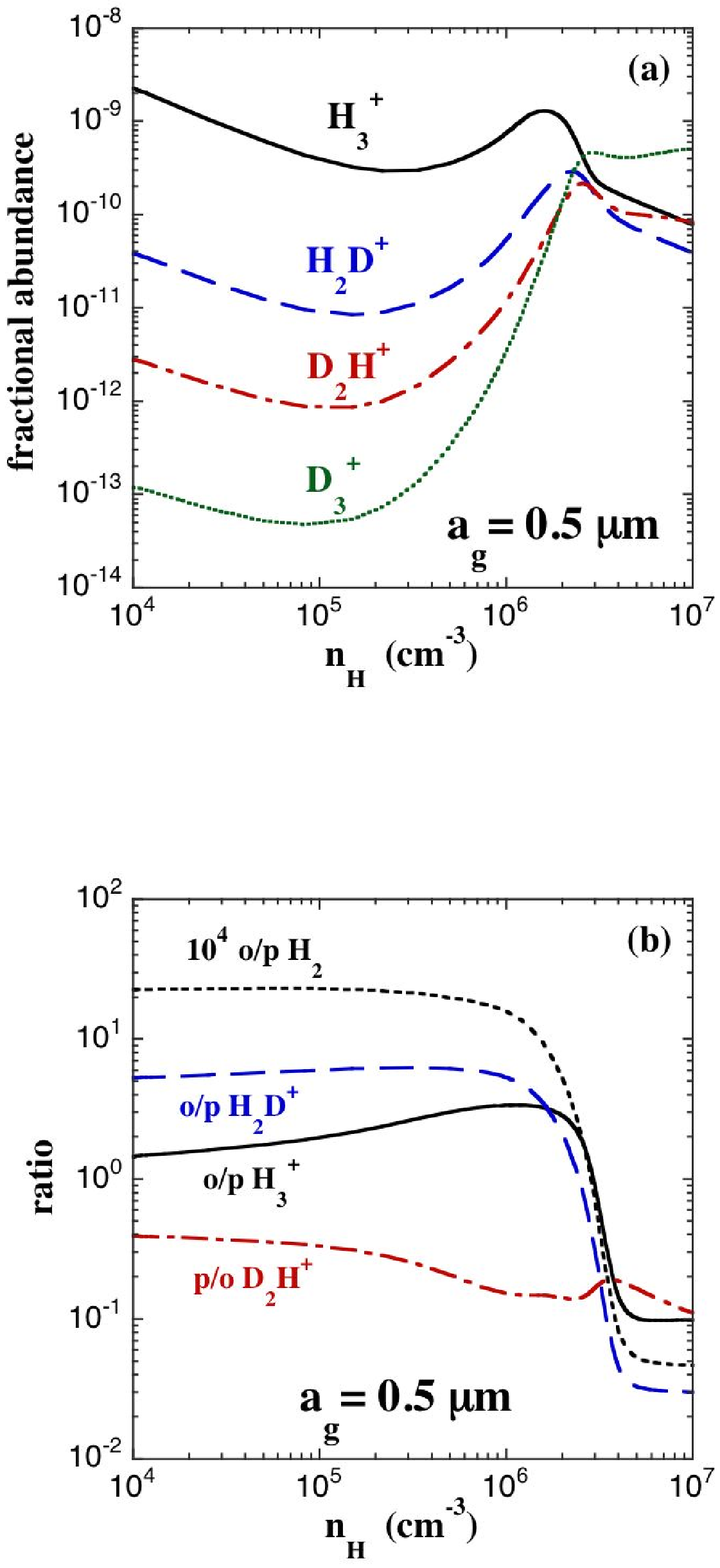}
\caption{(a) The fractional abundances of H$_3^+$ and its deuterated isotopes, and (b) the relative abundances of their ortho and para forms (in the sense of excited state divided by ground state abundance), for $a_{\rm g} = 0.5$~$\mu $m. The ortho:para H$_2$ ratio is also plotted, for reference.}
\label{H3+}
\end{figure}

The behaviour of H$_3^+$ and its deuterated forms is shown in Fig.~\ref{H3+}, for $a_{\rm g} = 0.5$~$\mu $m. The increase in the degree of deuteration of H$_3^+$, as freeze--out occurs, is apparent in this Figure; H$_2$D$^+$, D$_2$H$^+$ and D$_3^+$ are responsible for the deuteration of NH$_3$. The rapid decrease in the abundance ratios of the ortho and para forms of H$_3^+$ and H$_2$D$^+$ is seen, in Fig.~\ref{H3+}b, to coincide with the decrease in the ortho:para H$_2$ ratio; these ortho:para ratios are linked chemically, through proton--exchange reactions (cf. Flower et al. 2006).  The ortho:para H$_2$D$^+$ ratio depends on both $n_{\rm H}$ and $a_{\rm g}$. The maximum fractional abundance of ortho--H$_2$D$^+$ is attained for $n_{\rm H} \approx 10^6$~cm$^{-3}$ and is of the order of $10^{-10}$. The maximum in ortho--H$_2$D$^+$ occurs close to the maxima in the deuterated forms of NH$_3$, suggesting that observations of these species sample the same region of the core.
The fact that the ortho:para H$_2$D$^+$ ratio remains large for densities up to $n_{\rm H} \approx 10^6$~cm$^{-3}$ probably accounts, at least partly, for the ease with which ortho--H$_2$D$^+$ is being detected in prestellar cores. 

In Fig.~\ref{s-s2b}, we compare the predictions of our model with the observations of the degrees of deuteration of N$_2$H$^+$ and NH$_3$ in the prestellar core L1544. The observational and theoretical results are in satisfactory agreement for either of the values of $a_{\rm g}$ for which calculations are displayed, i.e. $a_{\rm g} = 0.5$~$\mu $m or $a_{\rm g} = 1.0$~$\mu $m. Grain sizes $a_{\rm g} < 0.5$~$\mu $m or $a_{\rm g} > 1.0$~$\mu $m would appear to be excluded by the observations, when interpreted in the context of our model.

\subsection{Relative population densities of nuclear spin states}

We note first that our expectations regarding the relative populations of the states $I = 0$, $I = 1$ and $I = 2$ of NH$_4^+$ (Appendix~\ref{Nions}) and the para:ortho NH$_3$ ratio (Appendix~\ref{NH3}) are confirmed by the results of the free--fall collapse model, which show these population ratios to be independent of the ortho:para H$_2$ ratio and of the gas density and equal to the numerical values derived in Appendix~\ref{AppA}.

As noted in Appendix~\ref{NH2D}, the only species for which more than one nuclear spin state has been observed in prestellar cores is NH$_2$D. The ortho:para NH$_2$D ratio which is predicted, on the basis of the arguments in Appendix~\ref{NH2D}, is 2:1; this may be compared with the value of 3:1 which is anticipated from the ratio of the degeneracies, $2I + 1$, of the corresponding nuclear spin states, $I = 1$ and $I = 0$. In practice, the observations reported by Shah \& Wootten (2001) are perhaps marginally more consistent with a ratio of 3:1 than 2:1; but, allowing for the associated observational error bars, the lower value (of 2:1) is consistent with their observations of 4 out of a total of 8 sources.

\section{Concluding remarks}
\label{Conclusions}

\begin{itemize}

\item We have investigated the chemistry of nitrogen--bearing species during the initial phases of protostellar collapse. We find that the timescales for these species to reach their steady--state, equilibrium values are large -- comparable to, or perhaps greater than, the lifetimes of the precursor molecular clouds. Even if, as we have supposed, steady state is attained prior to the onset of gravitational collapse, the nitrogen chemistry departs from steady state as soon as collapse begins. Large discrepancies develop, as $n_{\rm H}$ increases in the range $10^4 \le n_{\rm H} \le 10^7$~cm$^{-3}$, between the predictions of the free--fall model and the corresponding results in steady--state, for important nitrogen--bearing species, specifically N$_2$, NO, NH$_3$ and N$_2$H$^+$. The steady--state calculations underestimate the fractional abundances of N$_2$ and N$_2$H$^+$, and overestimate those of NO and NH$_3$, by factors which approach 2 to 3 orders of magnitude in the range of gas densities considered here.

\item We have assumed that the collapse occurs rapidly, on a timescale of approximately $4\times 10^5$ years. A longer evolutionary timescale would allow CO to freeze--out at lower densities than are observed. In addition, this timescale 
is compatible with current observational estimates of the lifetimes of prestellar cores. Another assumption of our model is that the initial chemical composition
corresponds to the gas phase being in steady state at the initial density. In fact, we have found that the chemical evolution predicted by our model is insensitive to this assumption, and, in particular, to the $n_{\rm C}/n_{\rm O}$ elemental abundance ratio and even to the extent to which elements heavier than helium have been incorporated into molecules and molecular ions. We have made ad hoc assumptions about the sticking coefficients, $S$, of atomic nitrogen and atomic oxygen. Clearly, laboratory work is needed to check the values of $S$(O) and $S$(N) which have been adopted.

\item We have considered the issue of the relative longevity of nitrogen--containing species: N$_2$H$^+$ and NH$_3$ are observed to be still present in the gas phase when other molecules, specifically carbon--containing species such as CO, have already frozen on to the grains. These observations suggest that a major reservoir of elemental nitrogen, probably N or N$_2$, has either a low adsorption energy or a low probability of sticking to grains. Recent experimental work (Bisschop et al. 2006) excludes these possibilities in the case of N$_2$. Accordingly, we have supposed that the sticking probability for atomic nitrogen, $S({\rm N}) < 1$. Because the chemical pathways leading to N$_2$H$^+$ and NH$_3$ are initiated by N(OH, H)NO [reaction (\ref{equ0.1}) above], and OH is produced from atomic oxygen, chemical considerations imply that $S({\rm O}) < 1$ also. Subject to these hypotheses, we find that the differential freeze--out of nitrogen-- and carbon--bearing species can be reproduced by our model of free--fall collapse when a sufficiently large grain radius ($a_{\rm g} \approx 0.5$~$\mu $m) is adopted. Although OH cannot be observed with sufficient angular resolution to investigate directly its chemical role in prestellar cores, NO can and has been observed, in L1544 and L183; these observations will be presented in a forthcoming publication (Akyilmaz et al. 2006, in preparation).

\item In effect, we have varied the values of the timescales characterizing the gas phase chemistry and depletion on to grain surfaces, relative to the timescale for free--fall collapse, investigating the consequences for observable quantities, such as the degree of deuteration of ammonia.  Our results indicate that $(n_{\rm g}/n_{\rm H})\pi a_{\rm g}^2$ is an order of magnitude lower than in the diffuse interstellar medium.  In a recent study of the thermal structure of B68, Bergin et al. (2006) reach a similar conclusion, for different, physical rather than chemical  reasons.  Of course, other approaches might be adopted, for example, to vary the cosmic ray ionization rate, which is a determinant of the timescale of the ion--neutral chemistry. However, test calculations have indicated that, unless the cosmic ray ionization rate is much larger than the value adopted here ($\zeta = 1\times 10^{-17}$ s$^{-1}$), it is not as important a parameter as $(n_{\rm g}/n_{\rm H})\pi a_{\rm g}^2$; and ionization rates $\zeta >> 10^{-17}$ s$^{-1}$ would give rise to gas kinetic temperatures which are higher than are observed.

\item The significance of the ortho:para H$_2$ ratio for the deuteration of species such as N$_2$H$^+$ and NH$_3$ (Flower et al. 2006) has been emphasized. Ortho--H$_2$ restricts the degree of deuteration of H$_3^+$, because the reverse of the reaction H$_3^+$(HD, H$_2$)H$_2$D$^+$, when it involves  ortho--H$_2$, is more energetically favourable and much more rapid at low temperatures than the corresponding reaction with para--H$_2$. Any reduction in the abundances of the deuterated species H$_2$D$^+$, D$_2$H$^+$ and D$_3^+$ leads to a fall in the abundances of species such as NH$_2$D, ND$_2$H and ND$_3$, which are produced in reactions with the deuterated forms of H$_3^+$. The timescale for the deuteration of N$_2$H$^+$, through reaction~(\ref{equ2}) and the analogous reactions with D$_2$H$^+$ and D$_3^+$, becomes comparable with the free--fall time. In the case of the prestellar core L1544, we have shown that the results of our model are consistent with the observed high levels of deuteration of N$_2$H$^+$ and NH$_3$ when $0.5 \lesssim a_{\rm g} \lesssim 1.0$~$\mu $m. Our previous study, of freeze--out and coagulation during protostellar collapse (Flower et al. 2005), indicates that such large grain sizes cannot be attained through coagulation {\it during} free--fall collapse; significant coagulation of the grains must have occurred {\it prior} to collapse.

\item Singly deuterated ammonia, NH$_2$D, was considered as a special case. It is special, in that it is the only species which has been observed in prestellar cores in both its ortho and para forms. As such, it provides a test of the statistical procedures which we have adopted to calculate the number densities of molecules in specific nuclear spin states. For NH$_2$D, we predict an ortho:para ratio of 2:1, which is consistent, to within the error bars, with results for 4 of the 8 prestellar cores observed by Shah \& Wootten (2001). The deuterated forms of ammonia attain their maximum fractional abundances for densities in the range $5\times 10^5 \lesssim n_{\rm H} \lesssim 5\times 10^6$ cm$^{-3}$. The maximum fractional abundance shifts to higher densities with increasing degree of deuteration and with increasing grain size. In particular, ND$_3$, is present with significant fractional abundance only in high density gas, where $n_{\rm H} \approx 4\times 10^6$ cm$^{-3}$. 

\end{itemize}

\appendix

\section{Nuclear spin statistics}
\label{AppA}

In the calculations reported above, the abundances not only of the chemical species but also of their individual nuclear spin states have been determined, where applicable. The cases of interest here involve molecules or molecular ions comprising two or more protons, such as H$_2$. Fermi--Dirac statistics require that the nuclear wave function should be asymmetric under exchange of identical protons, and this restriction associates states of a given nuclear spin symmetry with states of an appropriate rotational symmetry. In the case of H$_2$, states with total nuclear spin $I = 0$ (para--H$_2$) are associated with rotational states with even values of the rotational quantum number, $J$; states with $I = 1$ (ortho--H$_2$) are associated with odd $J$. Transitions between states of differing nuclear spin, $I = 0$ and $I = 1$ in this example, are induced by proton--exchange reactions with the ions H$^+$, H$_3^+$, H$_2$D$^+$, D$_2$H$^+$ and HCO$^+$. In view of the significance of H$_2$ in the chemistry of molecular clouds, it might be anticipated that its ortho:para ratio would influence the analogous ratios in other molecules and molecular ions.

In order to provide a framework for the interpretation of the numerical results which are presented in this paper, we establish the values of the relative populations of the nuclear spin states of a number of key species, in static equilibrium. The introduction of the distinction between ortho and para forms of molecules and molecular ions leads to a large increase in the number of chemical reactions which has to be considered. It is desirable but not feasible to consider each reaction in detail. Consequently, a simpler, statistical approach was adopted.

\subsection{NH$_2^+$, NH$_3^+$, NH$_4^+$}
\label{Nions}

Let us consider the nitrogen--bearing ions which are produced in the hydrogenation sequence which commences with N$^+$(ortho--H$_2$, H)NH$^+$; the corresponding reaction with para--H$_2$ is endoergic by almost 170~K and negligible at $T = 10$~K. NH$_2^+$ forms in the reaction of NH$^+$ with H$_2$

\begin{equation}
{\rm NH}^+ + {\rm H}_2 \rightarrow {\rm NH}_2^+ + {\rm H}
\label{equA1}
\end{equation}
for which the rate coefficient was taken to be $1.27\times 10^{-9}$ cm$^3$ s$^{-1}$. The rate coefficients for the reactions with para-- and ortho--H$_2$ individually were deduced on the basis of the following arguments. 

\begin{itemize}

\item When NH$^+$ reacts with para--H$_2$, in which the proton spins are ``anti--parallel'', the probability that it will pick up the proton with a spin parallel to that of its own proton is the same as the probability that it will pick up the proton with an anti-parallel spin. Thus, we adopted a rate coefficient of $6.35\times 10^{-10}$ cm$^3$ s$^{-1}$ for the production of both ortho-- and para--NH$_2^+$. 

In this case, a more rigorous argument, based on considerations of the density of states, supports the outcome of the simple approach. In the reaction

\begin{equation}
{\rm NH}^+ + {\rm H}_2({\rm p}) \rightarrow {\rm NH}_2^+({\rm p}) + {\rm H}
\label{equA2}
\end{equation}
the initial and final value of the total proton spin, $I$, of the reactants and of the products) is $I = 1/2$. On the other hand, in

\begin{equation}
{\rm NH}^+ + {\rm H}_2({\rm p}) \rightarrow {\rm NH}_2^+({\rm o}) + {\rm H}
\label{equA3}
\end{equation}
the final spin is $I = 1/2$ or $I = 3/2$. However, only those channels with $I = 1/2$ can contribute, because $I$ must be conserved. Thus, although the final density of states, $\sum _I(2I+1) = 6$ in this case, i.e. 3 times greater than in the reaction~(\ref{equA2}), which yields para--NH$_2^+$, only 2 of the 6 final states are available. Hence, the probability of forming ortho--NH$_2^+$ is the same as that of forming para--NH$_2^+$, as concluded above.

\item When NH$^+$ reacts with ortho--H$_2$, in which the proton spins are ``parallel'', half of the collisions will occur with the spin of the proton of NH$^+$ parallel to those of H$_2$, and half with the spin anti--parallel. Hence, the rate coefficients for producing ortho-- and para--H$_2$ are, once again, $6.35\times 10^{-10}$ cm$^3$ s$^{-1}$.

Now consider

\begin{equation}
{\rm NH}^+ + {\rm H}_2({\rm o}) \rightarrow {\rm NH}_2^+({\rm p}) + {\rm H}
\label{equA4}
\end{equation}
and

\begin{equation}
{\rm NH}^+ + {\rm H}_2({\rm o}) \rightarrow {\rm NH}_2^+({\rm o}) + {\rm H}
\label{equA5}
\end{equation}
In the first of these two reactions, $I = 1/2$ or $I = 3/2$ for the reactants, whereas $I = 1/2$ only for the products. In the second reaction, $I = 1/2$ or $I = 3/2$ for both the reactants and the products. One would conclude that the probability of forming ortho--NH$_2^+$ in (\ref{equA5}) is 3 times that of forming para--NH$_2^+$ in (\ref{equA4}).

It is instructive to pursue the reasoning somewhat further than considerations of  the density of states only. In the reaction~(\ref{equA5}), when $I = 1/2$, ortho--NH$_2^+$ may be considered to form by capture of N$^+$ by ortho--H$_2$, rather than capture of a proton by NH$^+$. If, in fact, N$^+$ capture is improbable for other (chemical) reasons, then the $I = 1/2$ channels are eliminated from ortho--NH$_2^+$ formation. One would then conclude that the probability of forming ortho--NH$_2^+$ is twice that of forming para--NH$_2^+$. Essentially, the reaction forming para--NH$_2^+$ goes through an intermediate complex of para--NH$_3^+$ ($I = 1/2$), whereas the reaction forming ortho--NH$_2^+$ goes through an intermediate complex of ortho--NH$_3^+$ ($I = 3/2$). However, as in the case of ortho-- and para--NH$_3$, whilst the ratio of the nuclear spin statistical weights is 2:1, there are twice as many rotational states of para as of ortho symmetry. In other words, the overall ratio of statistical weights is 1:1. Thus, one concludes finally that the probability of forming para--NH$_2^+$, through an intermediate complex of para--NH$_3^+$, is the same as the probability of forming ortho--NH$_2^+$, through an intermediate complex of ortho--NH$_3^+$.

\end{itemize}

Ortho-- and para--NH$_2^+$ are destroyed, at the same rate, by para--H$_2$, 

\begin{equation}
{\rm NH}_2^+({\rm o}) + {\rm H}_2({\rm p}) \rightarrow {\rm NH}_3^+({\rm o}) + {\rm H}
\label{equA6}
\end{equation}
\begin{equation}
{\rm NH}_2^+({\rm o}) + {\rm H}_2({\rm p}) \rightarrow {\rm NH}_3^+({\rm p}) + {\rm H}
\label{equA7}
\end{equation}
\begin{equation}
{\rm NH}_2^+({\rm p}) + {\rm H}_2({\rm p}) \rightarrow {\rm NH}_3^+({\rm p}) + {\rm H}
\label{equA8}
\end{equation}
and hence ortho-- and para--NH$_2^+$ are predicted to have the same abundance. The rate coefficients for reactions~(\ref{equA6}) and (\ref{equA7}) are taken to be equal to one half that for reaction~(\ref{equA8}), and so the overall branching ratio in favour of para--NH$_3^+$ is 3:1. Both para-- and ortho--NH$_3^+$ are removed by para--H$_2$, yielding NH$_4^+$ in the nuclear spin states $I = 0$, $I = 1$ or $I = 2$. Allowing for the 3:1 para:ortho NH$_3^+$ ratio, the $I = 0$ : $I = 1$ : $I = 2$ density ratios expected statistically are 3:4:1, when NH$_4^+$ forms.

According to Friedrich et al. (1981), the lowest $I = 0$, $I = 1$ and $I = 2$ of NH$_4^+$ are separated by only 148~mK in the ``librational ground state''. In this case, proton--exchange reactions with ortho--H$_2$ can interconvert the nuclear spin states at low temperatures, but the corresponding reactions with para--H$_2$ are endoergic (by 170.5~K) and are negligible at low $T$. The $I = 0$ : $I = 1$ : $I = 2$ ratios which are expected, in equilibrium, from proton exchange alone are, on the basis of statistical arguments, 3:4:1, i.e. the same as in the process of formation of NH$_4^+$, considered above. Thus, proton--exchange reactions would {\it not} be expected to modify the $I = 0$ : $I = 1$ : $I = 2$ ratios of NH$_4^+$. 

The experiments of Friedrich et al. (1981) were performed on NH$_4^+$ ions in a lattice, i.e. in the solid state, and their relevance to ions in the gas phase is not self-evident. In his book on {\it Infrared and Raman Spectra}, Herzberg (1945) discussed the case of CH$_4$, which is isoelectronic with NH$_4^+$. On the basis of his discussion, we should conclude that the nuclear spin state with the lowest energy, $I = 2$, is associated with the rotational state $J = 0$ (i.e. the ground rotational state), $I = 1$ with $J = 1$, and $I = 0$ with $J = 2$. Taking $E_J = B J(J + 1)$ and $B \approx 5$ cm$^{-1}$ (the approximate value of the rotational constant of CH$_4$; that of NH$_4^+$ should not be much different), all from Herzberg's text, gives an overall separation of the nuclear spin states $E(I = 0) - E(I = 2) \approx 43$~K. As this separation is much less than the difference in the energies of the ortho and para ground states of H$_2$ (170.5~K), it remains true that proton--exchange reactions with ortho--H$_2$ but not para--H$_2$ can interconvert the nuclear spin states of NH$_4^+$ at low temperatures. 

\subsection{NH$_3$}
\label{NH3}

Dissociative recombination of NH$_4^+$ with electrons yields both ortho--NH$_3$ ($I = 3/2$) and para--NH$_3$ ($I = 1/2$); the lowest state of para--NH$_3$ lies 23.4~K above the lowest state of ortho--NH$_3$. Allowing for the relative weightings of the population densities of the $I = 0$ : $I = 1$ : $I = 2$ states of NH$_4^+$ (see Appendix~\ref{Nions}), and the adopted values of the dissociative recombination coefficients, the expected para:ortho NH$_3$ ratio is 3:1. The para:ortho NH$_3$ ratio can be modified by proton-exchange reactions with H$_3^+$, H$_2$D$^+$ and D$_2$H$^+$; but these reactions are insufficiently rapid to significantly affect the ratio, even in the limit of static equilibrium (`steady--state'), when the computed value is close to 3.

NH$_4^+$, the precursor of NH$_3$, forms in reactions with {\it para}--H$_2$. On the other hand, NH$_4^+$ undergoes proton--exchange reactions with {\it ortho}--H$_2$. Consequently, we might expect the para:ortho NH$_3$ ratio to depend on the ortho:para H$_2$ ratio. However, as we have seen in Appendix~\ref{Nions}, the $I = 0$ : $I = 1$ : $I = 2$ population density ratios of NH$_4^+$ are predicted to be 3:4:1, independent of whether the formation or the proton--exchange reactions dominate. Consequently, the para:ortho NH$_3$ ratio is predicted to be independent of the ortho:para H$_2$ ratio.

\subsection{NH$_2$D}
\label{NH2D}

Finally in this Section, we consider the only species (NH$_2$D) whose ortho:para abundance ratio has been observed directly in prestellar cores (Shah \& Wootten 2001). Singly--deuterated ammonia, NH$_2$D, is produced by the dissociative recombination reaction  

\begin{equation}
{\rm NH}_3{\rm D}^+ + {\rm e}^- \rightarrow {\rm NH}_2{\rm D} + {\rm H}
\label{equA9}
\end{equation}
Its precursor, NH$_3$D$^+$, is formed principally in the deuteron--transfer reaction 

\begin{equation}
{\rm NH}_3 + {\rm H}_2{\rm D}^+ \rightarrow {\rm NH}_3{\rm D}^+ + {\rm H}_2
\label{equA10}
\end{equation}
and the analogous reactions with D$_2$H$^+$ and D$_3^+$. Accordingly, NH$_3$D$^+$ is formed with the same para:ortho ratio as NH$_3$, i.e. 3:1, where the lowest para state is energetically higher than the lowest ortho state. However, just as in the case of NH$_4^+$, proton--transfer reactions with ortho--H$_2$ modify the para:ortho ratio of NH$_3$D$^+$. Statistical considerations suggest that the rate coefficient for the reaction of ortho--H$_2$ with para--NH$_3$D$^+$, forming ortho--NH$_3$D$^+$, 

\begin{equation}
{\rm NH}_3{\rm D}^+({\rm p}) + {\rm H}_2({\rm o}) \rightarrow {\rm NH}_3{\rm D}^+({\rm o}) + {\rm H}_2({\rm p})
\label{equA11}
\end{equation}
is the same as for the reaction with ortho--NH$_3$D$^+$, forming para--NH$_3$D$^+$,

\begin{equation}
{\rm NH}_3{\rm D}^+({\rm o}) + {\rm H}_2({\rm o}) \rightarrow {\rm NH}_3{\rm D}^+({\rm p}) + {\rm H}_2({\rm p})
\label{equA12}
\end{equation}
Furthermore, these reactions are sufficiently rapid to determine the para:ortho NH$_3$D$^+$ ratio, which consequently becomes 1:1. Dissociative recombination of NH$_3$D$^+$ with electrons favours the production of ortho--NH$_2$D, relative to para--NH$_2$D, by a factor of 2. Thus, the ortho:para NH$_2$D ratio which is expected on the basis of these arguments is 2:1.

\begin{acknowledgements}

GdesF and DRF gratefully acknowledge support from the `Alliance' programme, in 2004 and 2005. We thank Paola Caselli for a critical reading of the original version of our paper.

\end{acknowledgements}

\end{document}